\documentclass[a4paper,fleqn,usenatbib]{mnras}

\usepackage[T1]{fontenc}
\usepackage{ae,aecompl}

\usepackage{graphicx}	
\usepackage{amsmath}	
\usepackage{amssymb}	

\usepackage{color}
\usepackage{natbib}
\usepackage{comment}

\title{The Winds from HL Tau}

   \title{The winds from HL Tau}

   \author[P.D. Klaassen et al.]{P. D. Klaassen,$^{1}$\thanks{pamela.klaassen@stfc.ac.uk}
          J. C. Mottram,$^{2,3}$ L. T. Maud,$^{2}$ and  A. Juhasz$^{4}$\\
$^{1}$UK Astronomy Technology Centre, Royal Observatory Edinburgh, Blackford Hill, Edinburgh EH9 3HJ, UK\\
$^{2}$Leiden Observatory, Leiden University, PO Box 9513, 2300 RA Leiden, The Netherlands\\
$^{3}$Max-Planck Institute for Astronomy, K\"{o}nigstuhl 17, 69117 Heidelberg, Germany\\
$^{4}$Institute of Astronomy, Madingley Road, Cambridge CB3 OHA, UK}

\date{Accepted 2016 April 25. Received 2016 April 25; in original form 2015 July 10}

\pubyear{2016}

\begin{document}
\label{firstpage}
\pagerange{\pageref{firstpage}--\pageref{lastpage}}
\maketitle

  \begin{abstract}
  Outflowing motions, whether a wind launched from the disk, a jet launched from the protostar, or the entrained molecular outflow, appear to be an ubiquitous feature of star formation.  These outwards motions have a number of root causes, and how they manifest is intricately linked to their environment as well as the process of star formation itself.

Using the ALMA Science Verification data of HL Tau, we investigate the high velocity molecular gas being removed from the system as a result of the star formation process.  We aim to place these motions in context with the optically detected jet, and the disk. With these high resolution ($\sim 1''$) ALMA observations of CO (J=1-0), we quantify the outwards motions of the molecular  gas.  We find evidence for a bipolar outwards flow, with an opening angle, as measured in the red-shifted lobe,  starting off at 90$^\circ$, and narrowing to 60$^\circ$ further from the disk, likely because of magnetic collimation.  Its outwards velocity, corrected for inclination angle is of order 2.4 km s$^{-1}$.  
\end{abstract}


\section{Introduction}

Protostars accrete mass through disks, and as material moves inwards through the disk, angular momentum builds up.  Without a mechanism to release this buildup of angular momentum, accretion would halt, as material reaches escape velocities.  There are a number of mechanisms for dispersing angular momentum, including the launching of winds, outflows and jets. They allow the remaining material to continue moving through the disk, eventually getting to the inner edge, and becoming available for accretion by the star. All three of these large scale phenomena are intrinsically linked with each other \citep[see, for example,][]{Panoglou12}. Jets are generally observed in high energy, ionised (optical) tracers and in radio continuum emission.  They are often associated with Herbig-Haro objects and show knotty or turbulent structures.   Highly collimated, high velocity jets are thought to entrain surrounding material forming molecular outflows, while winds are thought to be lifted directly from the disk, with the suggestion that the jets may be collimated by the wider angle MHD disk winds \citep[e.g.][]{Frank14}. These phenomena exhibit different properties from each other (e.g. in terms of collimation, gas velocities), however the terms `wind' and `outflow' are often used interchangeably since their observed properties are inherently similar (e.g. they can be detected in the same tracers).

Winds, being launched from the disk, come in two forms; photo-evaporative flows and molecular disk winds.  The former is caused by the forming star photoionising the disk surface  \citep[e.g.][]{Alexander06}, at which point highly energetic particles whose velocity exceeds the escape velocity will leave the system. These flows have velocities of $>$ 10 km/s and are seen in ionised atomic species.  The later, molecular disk winds, are launched from magnetic foot points within the disk, and are collimated by the magnetic field \citep[e.g.][]{Pudritz97}. They are a key mechanism for releasing angular momentum, and have been seen in molecular CO \citep[e.g.][]{Klaassen13a}.  Molecular disk winds can self-shield such that the wind gas remains cool and thus molecular \citep[e.g.][]{Panoglou12}.

Outflows, generally traced with molecular line emission in the (sub-)mm, highlight the interaction between a jet/wind and its environment: as the jet or wind pushes through the ambient material, it entrains a portion of that material (through processes such as turbulence and viscosity), accelerating it to higher velocities. Because outflows trace entrained material, their flow speeds tend to be lower, and their opening angles tend to be larger than those of jets or winds. We use the term `outflowing material' to describe the generalised molecular gas moving away from the disk, regardless of whether it is in a wind or an outflow.

HL Tau is an interesting region for testing the interconnectedness of these phenomena as both atomic jets \citep[e.g. detected in H$\alpha$,][]{Krist08} and molecular outflows \citep[e.g.][]{Lumbreras14} have previously been observed. HL Tau, at a distance of 140 pc \citep{Kwon11},  is a young protostar of between $0.55-1.3$ M$_\odot$ \citep{Stephens14,Brogan15}.

Using Near IR polarimetry measurements, \citet{Lucas04} found a strongly twisted magnetic field surrounding the disk in HL Tau. They attributed the twisting to the wind from the protostar.  They were unable to measure the field of the disk itself due to the high levels of extinction.  \citet{Stephens14} measured the magnetic field in the disk and found it to be dominated by a toroidal magnetic field component. They found that its structure is not consistent with either completely toroidal nor completely vertical structures;  the field morphology must come from a combination of these structures.

The collimated H$\alpha$ jet emission appears to be concentrated in the blue shifted lobe, which hosts a poorly collimated blue-shifted molecular outflow.   The [S{\sc ii}] emission from the jet is seen in both the red and blue lobes \citep{Mundt90}. \citet{Takami07}, showed that the 1.64 $\mu$m continuum emission of HL Tau exhibits a cross pattern with symmetric red and blue components on ~1" scales, suggesting that the outflow is symmetric across the disk axis. They also detect a collimated jet in [Fe {\sc ii}] distinct from this outflow.  \citet{Beck10} show that the Br$\gamma$ emission from HL tau is coincident with the blue shifted jet emission, however it is not as blue-shifted as the other jet tracers detected in this source to within their velocity resolution.

The blue jet coming from HL Tau (HH 151) was first detected in the early 1980s \citep{Mundt83}, and curiously: (1) only becomes bright in H$\alpha$ 20$''$ from the powering source, and (2) appears at an angle to its expected direction; its observable base is along the disk axis, however from there it proceeds away from the disk at an angle.  These two properties of the jet are generally ascribed to interaction with the wind from XZ Tau \citep[e.g.][]{Movsessian12}, because these two forming stars are indeed spatially co-located (not just projected to be close on the sky). The base of the elongated H$\alpha$ emission is coincident with the  [S{\sc ii}] jet emission \citep{Mundt90}, and the kink in the H$\alpha$ emission can be seen in their Figure 4.

The high velocity sub-mm emission in this region was most recently presented by \citet{Lumbreras14} at a resolution of $\sim$ 2$''$, showing collimated red- and blue-shifted outflowing material in CO. However, these two lobes are not aligned in a classical bi-polar fashion, with the blue shifted lobe offset from the line connecting the blue shifted (optical) jet, and red shifted emission. Lower resolution CO observations \citep{Welch00} show that this emission is at the edge of a `bubble' of blue shifted emission and likely interacting with XZ Tau which is located within, and powering, the bubble.  It should be noted that XZ Tau is  thought to be a triple system \citep{Carrasco09} with two of the components being M3 and M2 Classical T-Tauri stars  \citep{Krist08}.

In this paper we present an analysis of the ALMA Science Verification data of CO (J=1-0) taken as part of the long baseline campaign in 2014.  In Section \ref{sec:obs} we present a brief overview of the observations, in Section \ref{sec:results} we present the images of the largest scale CO emission recovered in these observations, highlighting the wind and outflow components. In Section \ref{sec:discussion} we analyse these results, de-project the wind velocities and constrain the wind launch radius.  In Section \ref{sec:conclusions}, we summarise our findings.

\section{Observations}
\label{sec:obs}

The observations presented here come from the ALMA archive, and were taken as part of the ALMA Science Verification, long baseline campaign (project 2011.0.00015.SV). The integration times, calibrators (and their purposes), observing date, and  minimum and maximum baseline lengths used in each of the executions contained in this set of observations are listed in Table \ref{tab:calibrators}, in which each execution is identified by the last four characters of the filename given in the ALMA archive. Details about the data, and reduction can be found in \citet{Brogan15}.

We used unmodified versions of the released reference images of the Band 3 continuum and CO emission. The released spectral line data were tapered to highlight the larger scale structures, which results in a larger synthesised beam. The beam sizes and rms noise limits for the datasets are given in Table \ref{tab:observations}.  Data were imaged and analysed in {\sc casa} \citep{CASA}. 

These observations were not designed to recover the large scale outflowing gas, thus much of the emission has been filtered out.  The shortest baselines were 15.5 m, which means the largest angular scale to which these observations could be sensitive is approximately 41$''$.  This corresponds to approximately 77\% of the band 3 primary beam (53$''$).

\begin{table*}
\caption{Calibrators and Observing Parameters}
\begin{tabular}{llllllll}
\hline \hline
&Time on & \multicolumn{3}{c}{Calibrators} & & \multicolumn{2}{c}{Baseline Lengths}\\ \cline{3-5} \cline{7-8}
Execution  & Science Source & Phase & Bandpass & Flux & Observing Date & Max & Min\\
 & (min) &&&&(m) &(m)\\
 \hline
 X5fa & 30 & J0431+2037 & J0510+1800 & J0510+1800 & 2014-10-28 & 15238 & 15.5\\
X845 & 30 & J0431+2037 & J0423-0120 & J0510+1800 & 2014-10-28 & 15238 & 15.5\\
X3b2 & 30 & J0431+2037 & J0510+1800 & J0510+1800 & 2014-10-28 & 15238 & 15.5\\
X220 & 20 & J0431+2037 & J0423-0120 & J0423-0120 & 2014-11-11 & 15238 & 15.5\\
X693 & 21 & J0431+2037 & J0423-0120 & J0510+1800 & 2014-11-13 & 15238 & 15.2\\
X5b2 & 30 & J0431+2037 & J0510+1800 & J0423-0120 & 2014-11-11 & 15238 & 15.5\\
X306 & 30 & J0431+2037 & J0423-0120 & J0423-0120 & 2014-11-14 & 15238 & 15.2 \\
\hline \hline 
\end{tabular}
\label{tab:calibrators}
\end{table*}

\begin{table}
\caption{Properties of the ALMA observations used in this study. }
\begin{tabular}{lrr@{$\times$}lrr}
\hline
Species & Frequency & \multicolumn{3}{c}{Synthesised beam}&{rms noise}\\
(incl.  &  & B$_{maj}$ & B$_{min}$ & B$_{pa}$ &\\
transition) & (GHz)& ($''$) & ($''$) & ($^\circ$) &\\
\hline
CO (J=1-0)& 115.271 &0.98&0.90&-9&8.1 \\
Continuum & 115.136 & 0.09&0.06& -179 &0.01\\
\hline
\end{tabular}
{{\bf Note:} rms noise levels have units of mJy/beam per  1 km s$^{-1}$ channel  for the CO line emission, and mJy/beam for the continuum.}
\label{tab:observations}
\end{table}

Comparing our CO observations to those of \citet{Welch00} from BIMA at a resolution of $\sim$ 7$''$ shows that much of the CO emission in the blue-shifted emission is likely filtered out of our observations; their structures are much larger ($\sim$ 2.5$'$) than those seen with these ALMA observations. In terms of the red-shifted emission lobe, the size and shape of their recovered structures (see their Figure 2) are consistent with those seen with  ALMA. Because their observations are in $^{13}$CO (J=1-0) there are too many uncertainties (including an isotope ratio) to quantify the amount of missing flux in our $^{12}$CO (J=1-0) observations.

\subsection{ Other Archival Data}

In addition to the ALMA Science Verification data,  we make use of archival HST ACS data for this region, showing the H$\alpha$ and [N{\sc ii}] image first presented in \citet{Krist08}. 

\section{Results}
\label{sec:results}
 
Here we present the outflowing molecular gas properties, highlighting the morphological properties of the red (Section \ref{sec:red_wind}) and blue (Section \ref{sec:blue_wind}) emission and the opening angle of the red component of the flow (Section \ref{sec:opening}).  We further calculate the flow kinematics inferred from the emission and its velocity structure, making use of the known inclination angle on the sky (Section \ref{sec:kinematics}).

For our analysis, we have made extensive use of integrated intensity (zeroth moment), and intensity weighted velocity (first moment) maps of the CO emission.  These maps were clipped at 3 and 5 $\sigma$ respectively, using the noise level quoted in Table \ref{tab:observations} and were integrated over the velocity ranges of 0-5.5 km s$^{-1}$ and 7-20 km s$^{-1}$ respectively for the blue and red shifted emission.  We note that the spatial extent of the red-shifted emission detected in these observations is consistent with that of \citet{Lumbreras14} observed with the SMA.  To clarify, what \citeauthor{Lumbreras14} call a `wide-angle outflow', we discuss as an `outflow entrained by a wide-angle wind' (which we shorten to `flow'), to be more consistent with the terminology in \citet{Arce07}.

\begin{figure}
 \includegraphics[width=0.5\textwidth]{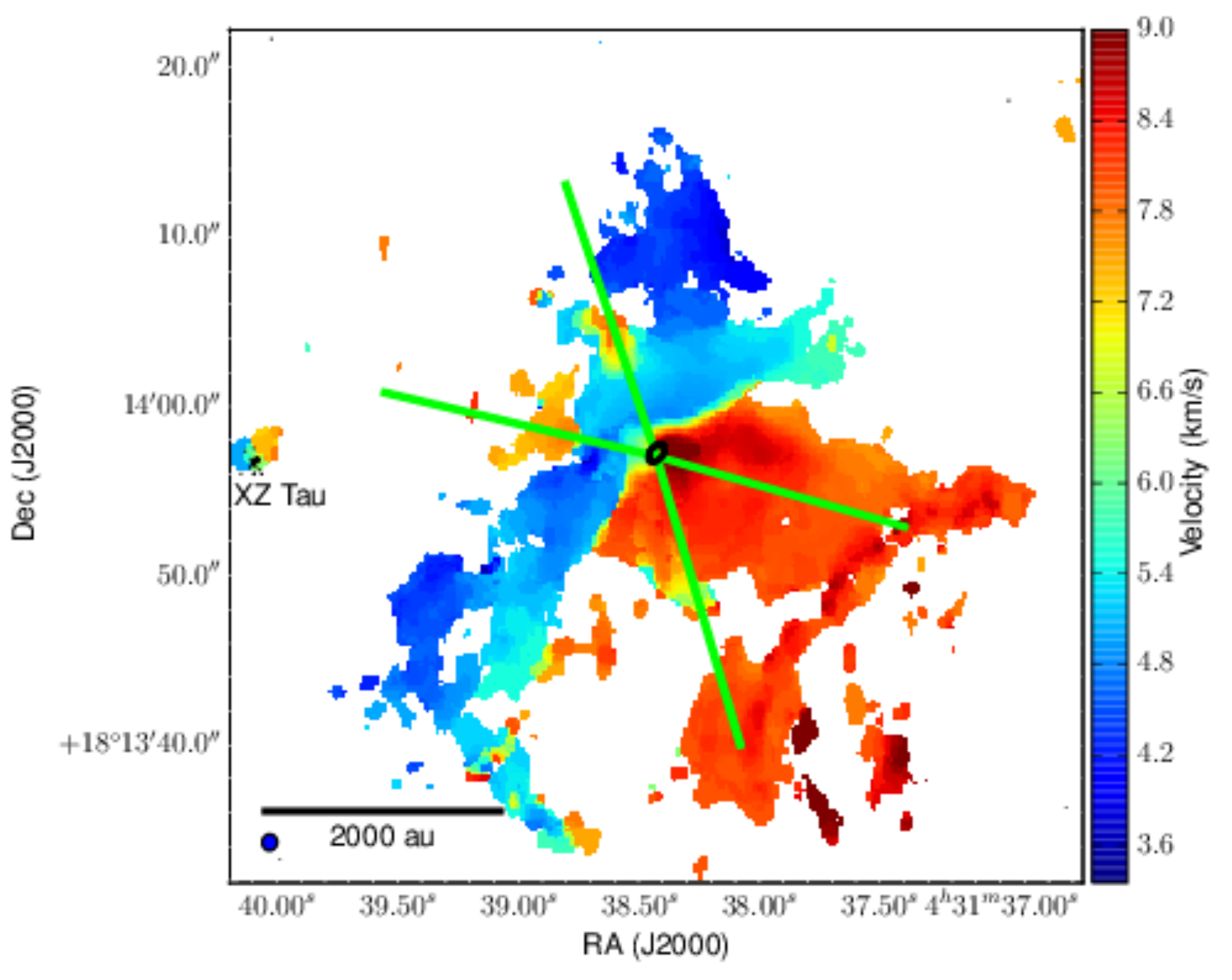}
 \caption{First moment map of the CO emission from HL Tau (cut at 10 $\sigma$, integrating over 0 to 20 km s$^{-1}$).  The black contours show the 5 to 25 $\sigma$ levels of the continuum emission to highlight the position of the disk.  Note that the emission (CO and continuum) at the left edge of the map is that from XZ Tau, as indicated. The green lines through the red lobe correspond to the white lines in Figure \ref{fig:red-wind}, which are then transposed onto the blue lobe. }
 \label{fig:CO_mom1}
 \end{figure}

  Figure \ref{fig:CO_mom1} shows the first moment map of the CO emission, highlighting the positions of the red and blue shifted emission lobes. The colourscale shows the velocities of the outflowing material in CO. This map already shows the complex morphology of the CO emission. It highlights the sharp transition between red and blue emission, and that red shifted emission is present towards the outer edge of the blue shifted emission (away from the disk).  As highlighted by green lines, these two red features make an angle with the disk quite similar to the opening angle of the red flow, as discussed further below.

 \subsection{Red-shifted emission}
 \label{sec:red_wind}

  \begin{figure}
 \includegraphics[width=0.5\textwidth]{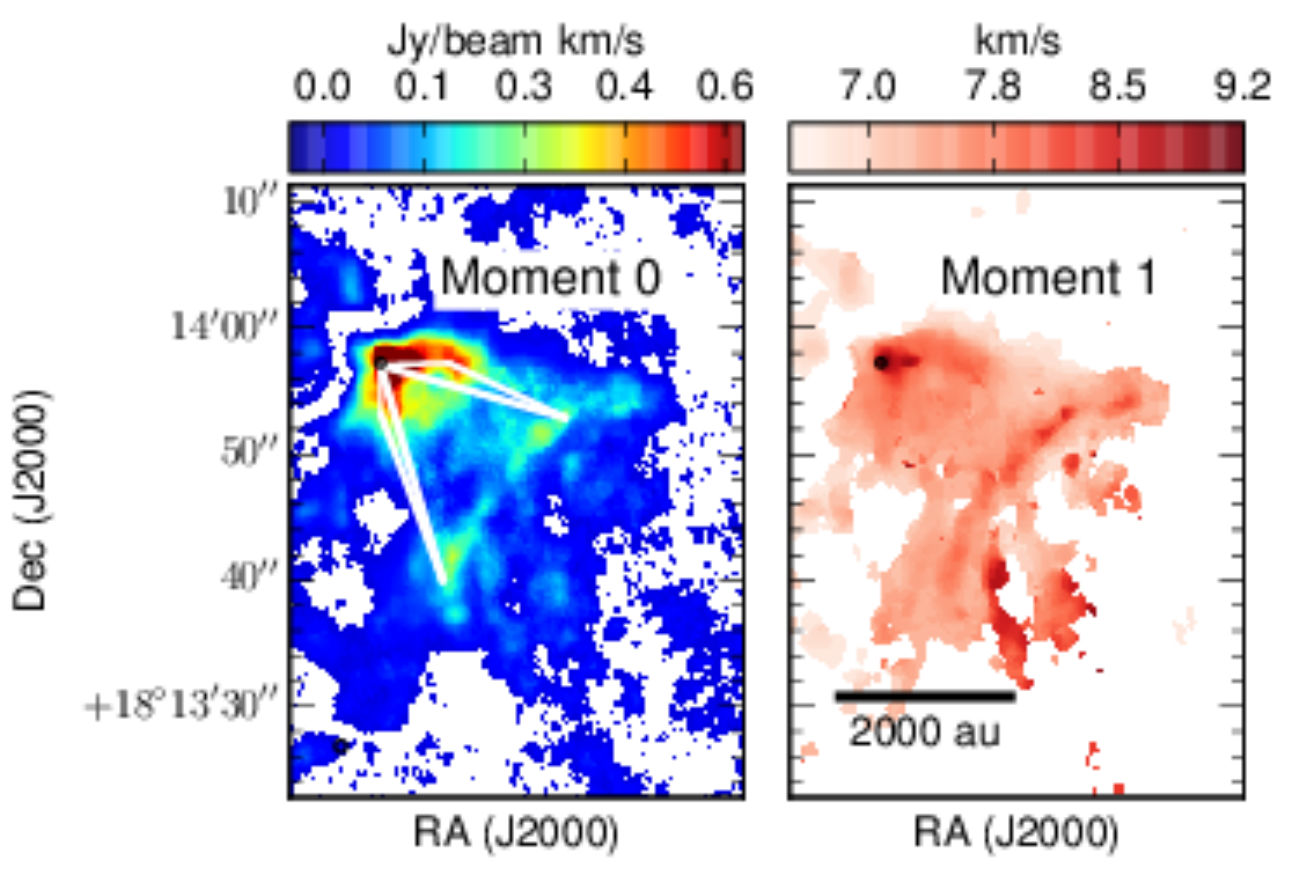}
 \caption{Red-shifted CO moment maps integrated over  7-20 km s$^{-1}$. Note that in the integrated intensity map (left panel), the edges of the outflow appear to be limb brightened. These edges are what we use to calculate the opening angle of the flow, and how its collimation factor varies along the flow. The white lines indicate the opening angles of the hourglass red-shifted flow.  Close to HL Tau, the opening angle is $\sim 90^\circ$, however, further out, the flow appears to have an opening angle closer to $60^\circ$.}
 \label{fig:red-wind}
 \end{figure}

 Because the red-shifted lobe appears to be limb brightened (the edges of the outflow appear to have the greatest intensities, see Figure \ref{fig:red-wind}), it is likely that the red emission is primarily coming from the conical edges of a flow, possibly with an excavated central cavity. The red outflowing material has an `hourglass' morphology which could be suggestive of magnetic collimation \citep[see, for instance][]{Stephens13}.  The previous observations of \citet{Lumbreras14} were of too low a resolution to see this limb brightening.

 \subsection{Blue-shifted emission}
 \label{sec:blue_wind}
 
\begin{figure}
 \includegraphics[width=0.5\textwidth]{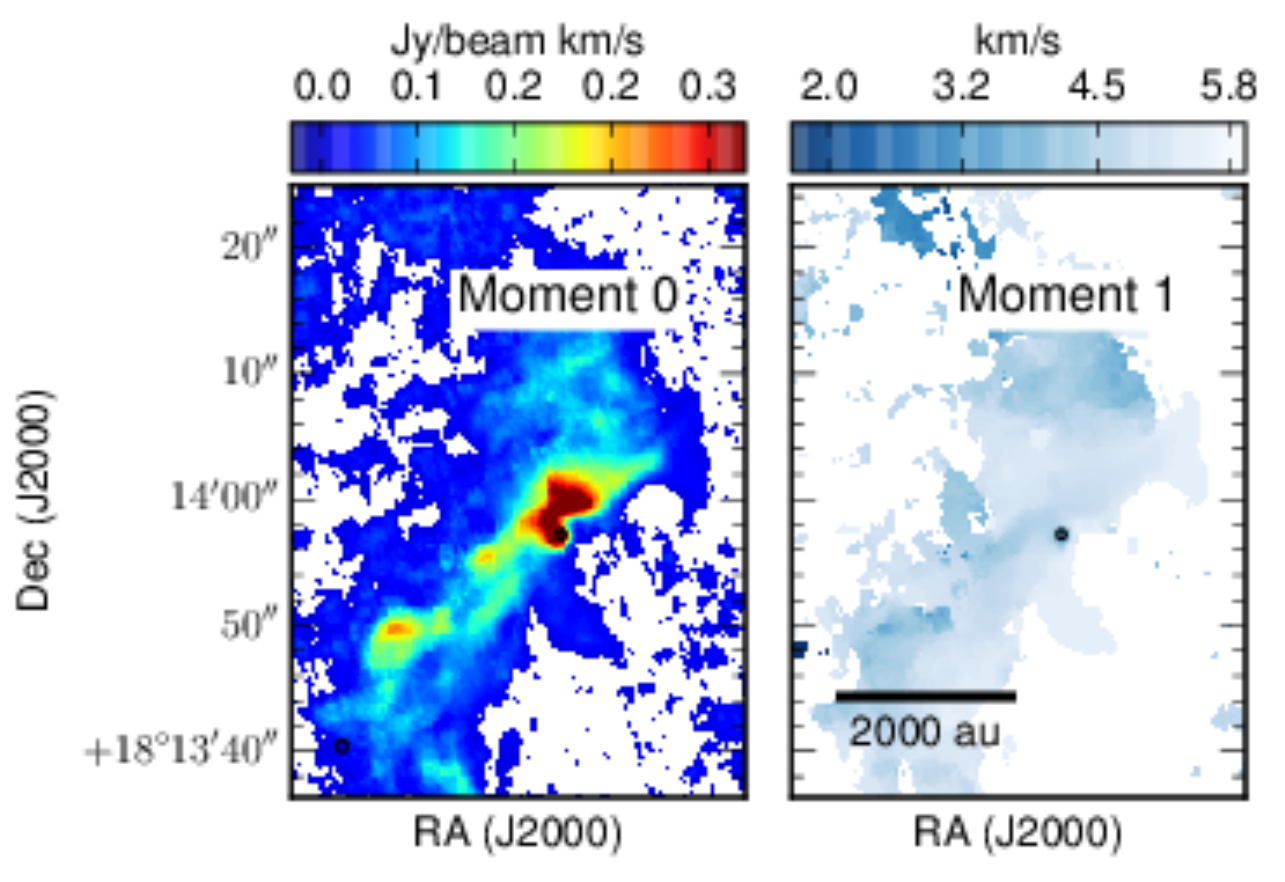}
 \caption{Blue-shifted CO moment maps integrated over  0-5.5 km s$^{-1}$. Note that both the integrated intensity and intensity weighted velocity increase significantly towards the centre (northern edge) of the blue shifted emission.  This is likely due to interaction with the jet (not plotted).}
 \label{fig:blue-wind}
 \end{figure}

\begin{figure}
 \includegraphics[width=0.5\textwidth]{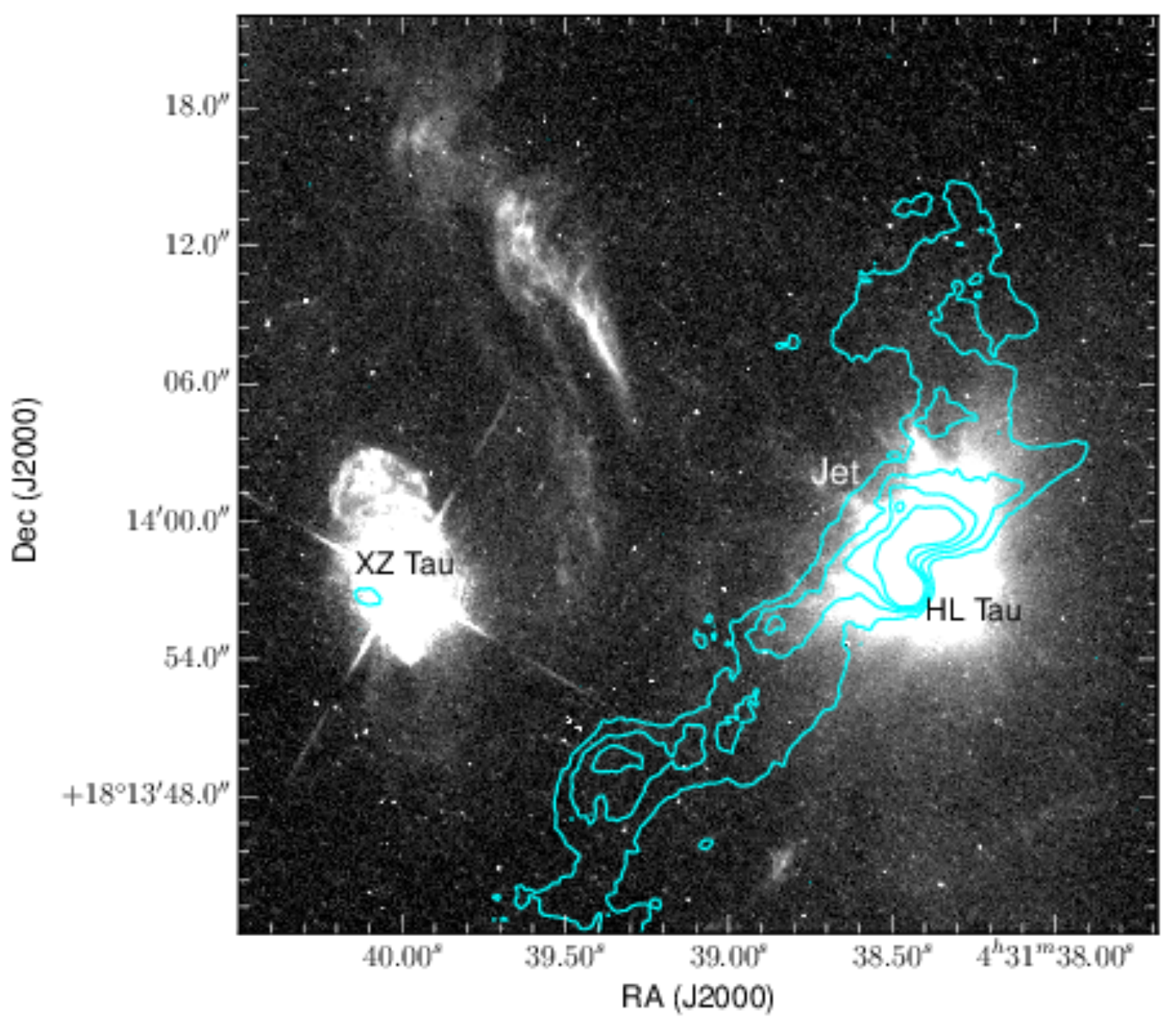}
 \caption{Blue shifted wind (blue contours, starting at 15\% of the peak intensity, increasing in levels of 10\%), overlaid on HST ACS greyscale image of a combination of H$\alpha$ and [N{\sc ii}].}
 \label{fig:blue_outflow}
 \end{figure}

  Figure \ref{fig:blue-wind} shows the moment maps of the blue-shifted CO emission.   The brightest blue-shifted emission comes from quite near the disk, where the H$\alpha$ emission is strongest as well. As can be seen in Figure \ref{fig:blue_outflow}, towards the centre of the flow axis (which arises between the two peaks in the blue shifted material, and is labelled as `Jet'), there is a small jet like structure protruding from the H$\alpha$ emission.
 
 The blue-shifted emission is not as collimated as the red, probably due to a combination of effects including the interaction with the XZ Tau bubble \citep[which is expanding within the blue shifted CO emission from HL Tau, see for instance][]{Welch00}.

 \subsection{Opening angle}
 \label{sec:opening}

The left panel of Figure \ref{fig:red-wind} shows the opening angle of the red-shifted fow. Its  morphology can be decomposed into two components: one with a small scale opening angle of 90$^\circ$, and a larger scale (collimated) opening angle of of 60$^\circ$. This hourglass morphology  could be due to magnetic collimation, like is the case for L1157-mm \citep{Stephens13}.   The magnetic field directions found by \citet{Lucas04} and \citet{Stephens14} are suggestive of field orientations which could collimate the fow. Under the assumption that we are not observing a special orientation on the sky, we expect this limb brightened morphology to be representative of a conical morphology in 3D.   

Additionally, we note that the 60$^\circ$ lines from Figure \ref{fig:red-wind} are drawn in Figures \ref{fig:CO_mom1} (in green) and Figure \ref{fig:CO_contamination} (in white) to help guide the eye in these plots. Note that in these two figures, the lines have been transposed onto the blue emission as well.

 \begin{figure}
 \includegraphics[width=0.5\textwidth]{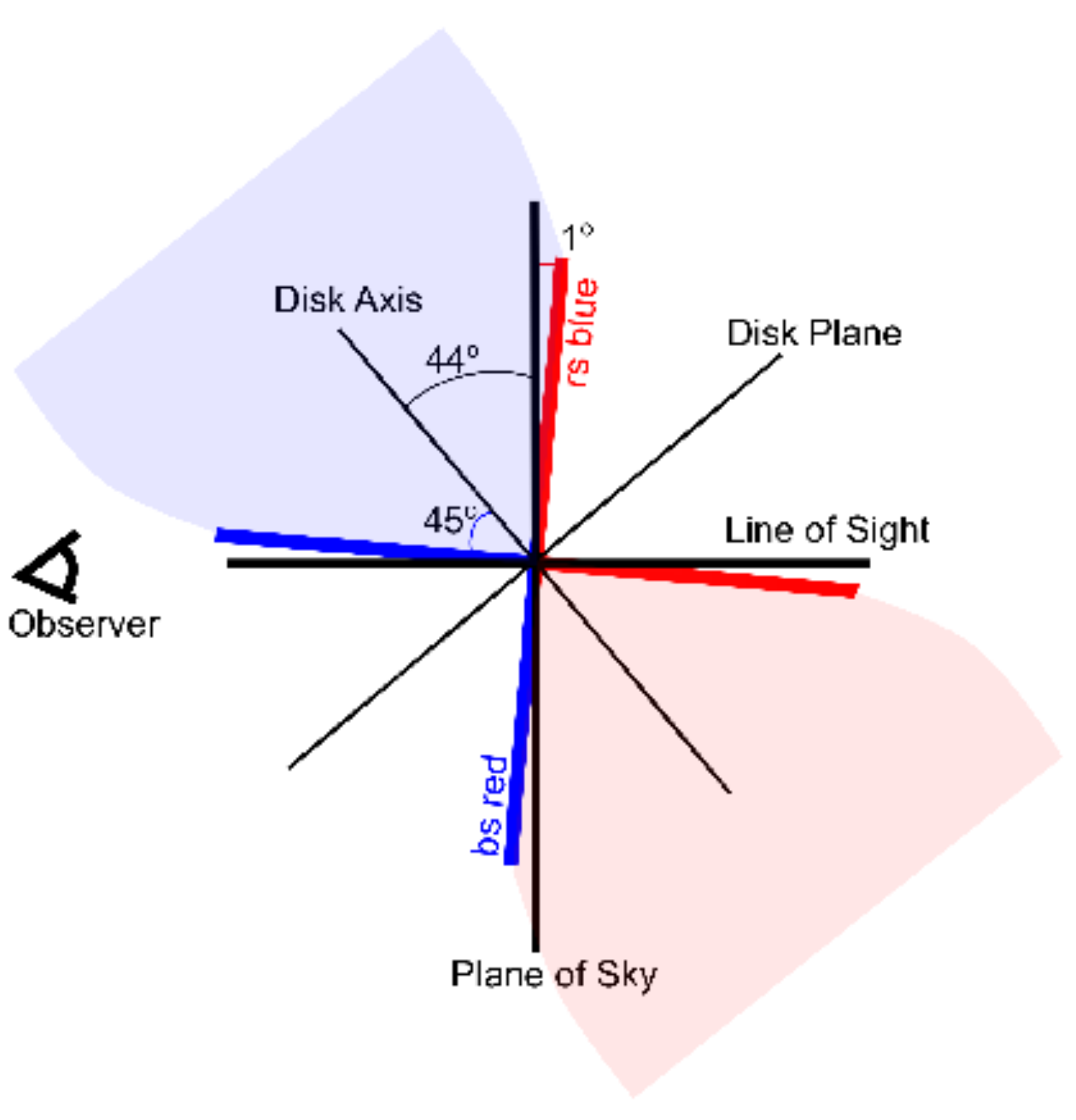}
 \caption{Cartoon of the relationship between the fow, its 90$^\circ$ opening angle at the base, and the inclination angle of the disk (which is assumed to be perpendicular to the outflow).  This representation explains why there is blue-shifted emission coincident with the red fow, and red-shifted emission coincident with the blue fow. }
 \label{fig:wind_cartoon}
 \end{figure}

 \begin{figure}
 \includegraphics[width=0.5\textwidth]{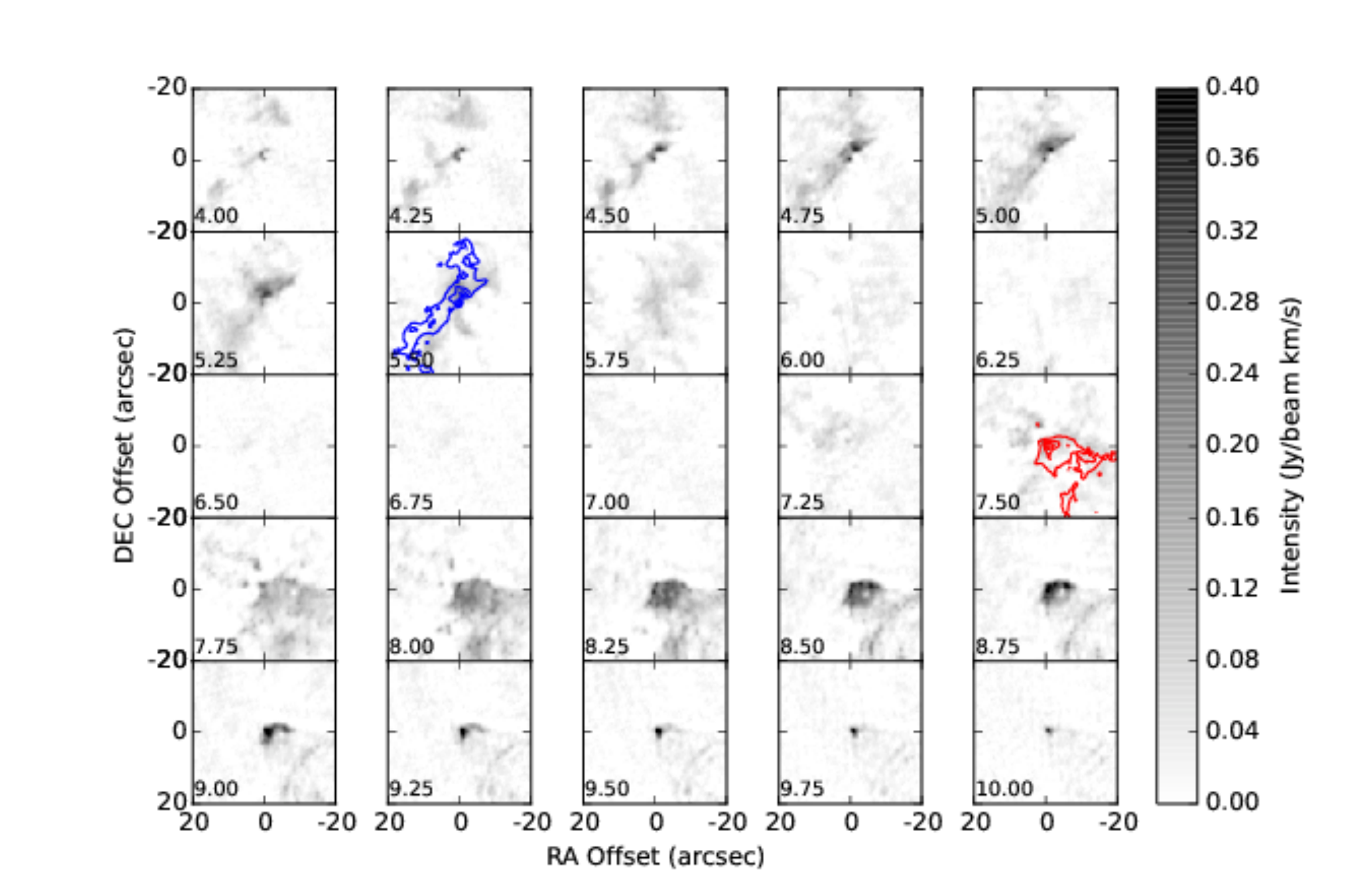}
 \caption{Channel map of the CO emission from 4 to 10 km s$^{-1}$. Note that there is very little emission detected at the systemic velocity ($\sim$ 6.5 km s$^{-1}$), which may be due to filtering. The integrated intensities of the blue and red shifted emission  are shown as contours (from 10\% to 70\% of the peak intensity) in the two channels at 5.5 and 7.5 km s$^{-1}$, where the counter-flow emission is strongest. The intensity contours are taken over the same velocity intervals as the moment 0 maps shown in Figures \ref{fig:red-wind} and \ref{fig:blue-wind} }
 \label{fig:CO_channel}
 \end{figure}

  \subsection{Flow Energetics}
 \label{sec:kinematics}

From the line wing CO emission, we quantified the mass, momentum and energy in the fow.  The intensity in each velocity channel was summed, and/or multiplied by the velocity of that channel to determine the mass (M $\propto \int Td$v), momentum (P $\propto$ M*v) and  energy (E $\propto M*v^2/2$). Here we assume the CO is optically thin, and that there is likely missing flux from the largest scale structures in these data, therefore our column density and mass estimates are lower limits. Using the extent of the red flow (20$''$), and the maximum velocity of the red-shifted emission (5 km s$^{-1}$), we estimate a kinematic age of approximately 2600 yr, with which we quantify the mechanical luminosity (L = E/t), and mass loss rate ($\dot{M}$ = M/t) of the flow.  We note that this is an extreme lower limit to the age of the flow. The constants of proportionality in the first few equations are related to the abundance of CO (taken to be 1$\times10^{-4}$), the ambient temperature (T= 50 K), and a multiplicative constant for CO related to the energy of the J=1-0 transition, the partition function, and the degeneracy of the state (see Appendix \ref{sec:outflow_calculations}).  We note that since the inclination angle of the flow is known (See Section \ref{sec:contamination}), the velocities used in the calculation have been corrected for inclination, and the results are given in Table \ref{tab:energetics1}.   Changing our assumed temperature by 20 K up or down changes our mass estimates by $\sim 35\%$. We note that our derived outflowing gas mass is approximately half that found in \citet{Lumbreras14}, while our momentum and kinetic energies are $<$ 10\% of their values.  The mass comparison gives an estimate of our missing flux, since we used the same LTE (T=50K) assumption.  Without knowing their velocity integration methods or limits, we cannot comment on the discrepancies between the momentum and energy calculations.

\begin{table}
\caption{Derived Flow Kinematics (assuming a kinematic age of 2650 yr)}
\begin{tabular}{llr@{$\pm$}lr@{$\pm$}l}
\hline\hline
Quantity & Units & \multicolumn{2}{c}{Red}& \multicolumn{2}{c}{Blue}\\
\hline
Mass &($\times 10^{-4}$ M$_\odot$) & 0.73 & 0.00 & 11.14 & 0.00\\
Momentum &($\times 10^{-4}$ M$_\odot$ km/s) & 2.21 & 0.05 & 14.77 & 0.51\\
Energy &($\times 10^{40}$ erg) & 1.52 & 0.05 & 4.74 & 0.25\\
Luminosity &($\times 10^{-5}$ L$_\odot$) & 4.72 & 0.15 & 14.70 & 0.76\\
Mass Loss &($\times 10^{-8}$ M$_\odot$/yr) & 2.73 & 0.00 & 41.96 & 0.00\\
\hline
\multicolumn{2}{c}{Velocity Integration Limits} & \multicolumn{2}{c}{9$\sim$ 20} & \multicolumn{2}{c}{ -4 $\sim$ 6}\\
\hline
\end{tabular}
\label{tab:energetics1}
{{\bf Note:} The uncertainties on the Mass and Mass Outflow rates are smaller than the precisions listed here. The method for calculating these uncertainties is presented in Section \ref{sec:outflow_calculations}, which takes into account the rms noise of the observations, but not things like CO opacity or abundance and ambient temperature assumptions.}
\end{table}

\section{Discussion}
\label{sec:discussion}


\subsection{fow  inclination angle}
\label{sec:contamination}

As discussed in Section \ref{sec:opening}, and shown in the left panel of Figure \ref{fig:red-wind}, the fow has an opening angle of $\sim90^\circ$ at its base. We analyse this opening angle using the assumption of conical symmetry about the fow axis. This assumption is supported by the observation that the fow is limb brightened (left panel of Figure \ref{fig:red-wind}).

\citet{Brogan15} fit the $uv$ plane visibilities of the dust continuum emission, and found an inclination angle of $46\pm0.2^\circ$ for the disk in the plane of the sky. If the fow axis is perpendicular to the disk axis, then it should have an inclination angle of 44$^\circ$ with respect to the line of sight. This, coupled with the fow opening angle of 90$^\circ$ suggests part of each wind lobe should exhibit some `counter-flow' emission, i.e. emission within a certain lobe that, due to the large opening angle, has crossed the plane of the sky and therefore appears to be flowing in the opposite direction.  As shown in the cartoon of Figure \ref{fig:wind_cartoon}, there should be a small portion of the blue fow which has red velocities - labelled as `rs (red-shifted) blue'. Similarly for the red fow, there is a small component labelled `bs (blue-shifted) red'. These `counter-flows', because they are close the plane of the sky, will have very small line of sight velocities.

\begin{figure}
 \includegraphics[width=0.5\textwidth]{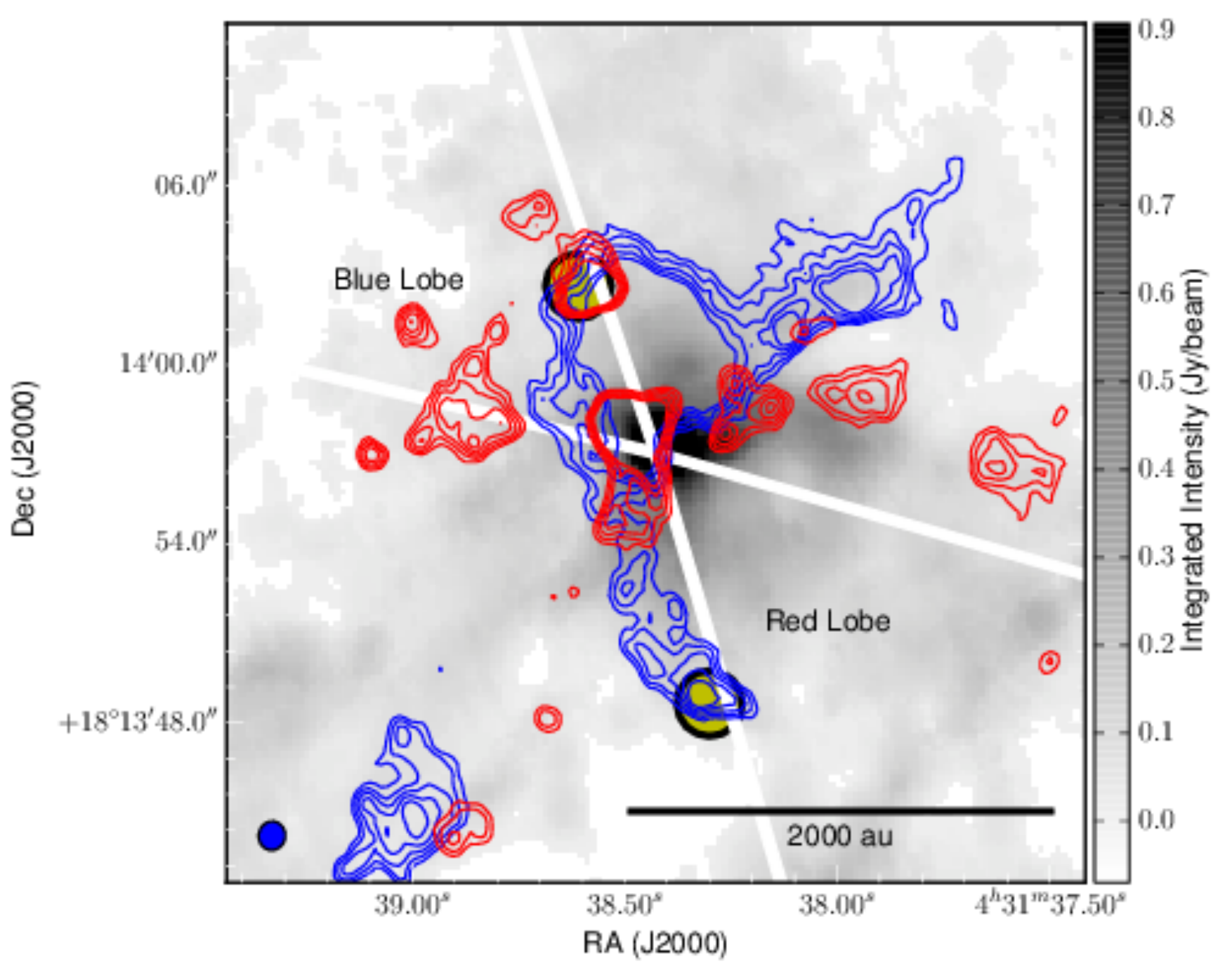}
 \caption{Overlap in red and blue shifted emission in the respectively `other' side of the flow.  The red contours consist of red-shifted emission at 7.5 km s$^{-1}$, while the blue contours show blue-shifted emission at 5.5 km s$^{-1}$. The labels show the predominant flow velocity within each cone; opposite to the low velocity contamination shown here.  The contours cover 10-15 $\sigma$ in 1 $\sigma$ steps.  Note that the bulk of the outflow material is not captured in these two channels, but exists at higher velocities from systemic (6-7 km s$^{-1}$). The while lines correspond to the green ones in Figure \ref{fig:CO_mom1}, and cross at the position of the disk.  The background greyscale shows the integrated intensity of the CO emission taken over the velocity range of the channel map in Figure \ref{fig:CO_channel}. The two ellipses used in Figure \ref{fig:CO_spectra} come from regions of high `contamination' of red/blue emission within the two yellow circles shown here.}
 \label{fig:CO_contamination}
 \end{figure}

\begin{figure}
 \includegraphics[width=0.5\textwidth]{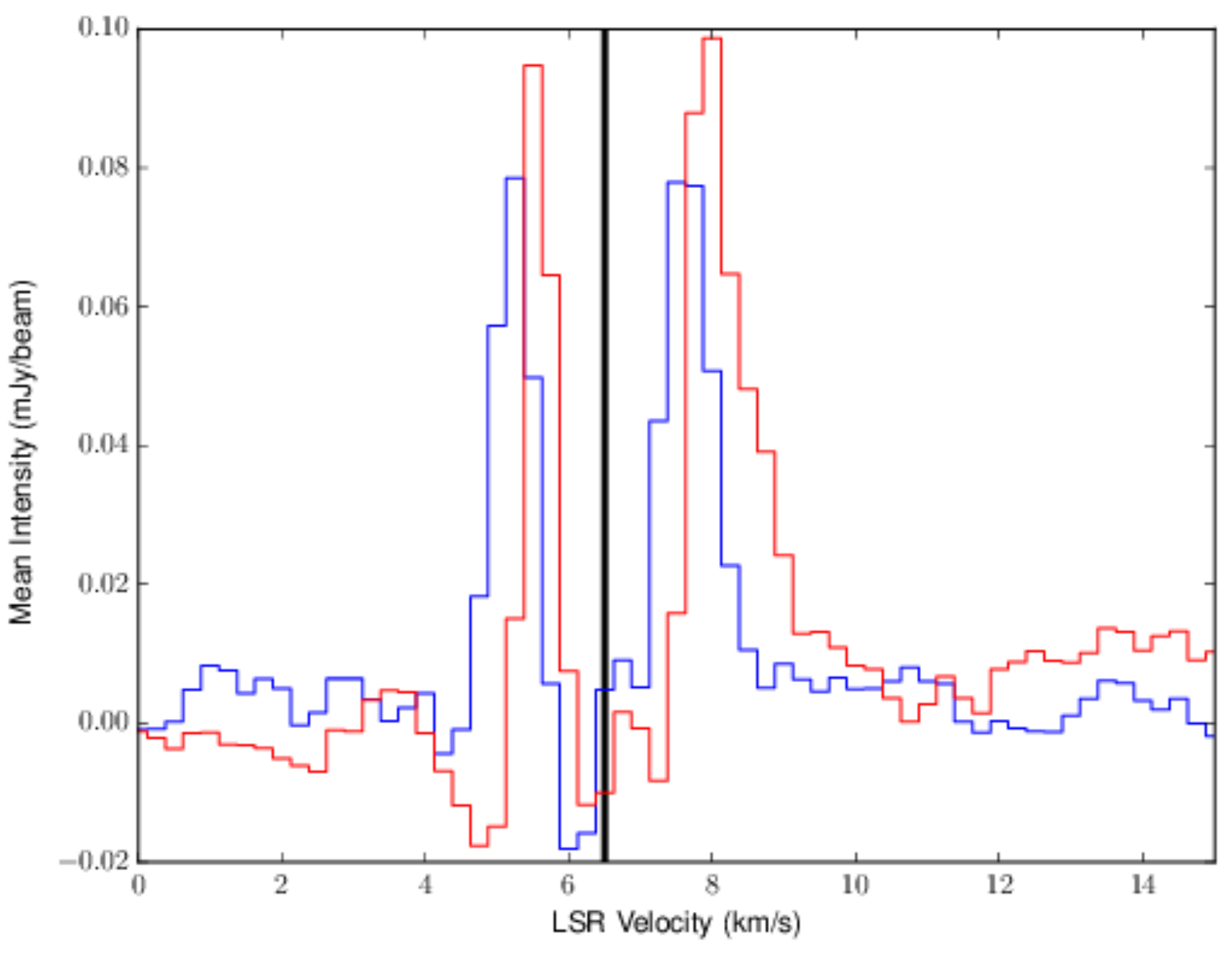}
 \caption{Spectra taken from two regions within the blue-shifted (blue line) and red-shifted (red line) portions of the wind where contamination from gas at the opposite velocities is strongest. The black vertical line represents the LSR velocity (6.5 km s$^{-1}$) suggested by \citet{Brogan15}.}
 \label{fig:CO_spectra}
 \end{figure}

These `rs blue' and `bs red' counter flows are highlighted in the channel maps of Figure \ref{fig:CO_channel} in the 5.5 and 7.5 km s$^{-1}$ channels. In these two channels, the integrated emission of the blue and red shifted winds are overplotted (respectively) with blue and red contours. This is to demonstrate where the blue and red wind lobes are, and highlight that there is additional emission in these channels spatially co-incident with the opposite wind lobe.  The `counter-flow' emission components at 5.5 and 7.5 km s$^{-1}$ are also shown in Figure \ref{fig:CO_contamination} as blue and red contours (respectively). Figure \ref{fig:CO_spectra} shows the CO spectra within the two black circles in Figure \ref{fig:CO_contamination} to highlight that there is significant `counter-flow' emission at these positions. Note that these circles were chosen a few arcsec from the disk to avoid contamination from the rotationally symmetric disk emission.

This shows that there is `rs blue' and `bs red' emission in this fow, which confirms that the opening angle is large, and that the fow is indeed perpendicular to the disk, as expected.

With the inclination angle known (with respect to the plane of the sky), the line of sight velocities of the red and blue emission known, and the opening angle of the cone assumed to be circularly symmetric (i.e. the opening angle measured in Figure \ref{fig:red-wind} is consistent along the line of sight as well), we can estimate the true velocities of the red and blue flows solving:

\begin{equation}
 \cos(1^\circ)*V_{blue} = V_b -V_{LSR}
 \end{equation}
 \begin{equation}
 \sin(1^\circ)*V_{blue} = V_r - V_{LSR} 
 \end{equation}

\noindent where $V_{blue}$ is the average velocity of the blue flow, $V_{LSR}$ is the LSR velocity $V_r$ and $V_b$ are the peak velocities red and blue shifted emission of the two components shown in the spectrum.  Solving Equation 1 for $V_{LSR}$ and putting that into Equation 2, the only unknown becomes $V_{blue}$, since $V_r$ and $V_b$ can be read from Figure \ref{fig:CO_spectra}.  The same method was used to calculate the red-shifted fow velocity. From these analyses, we find red and blue shifted wind velocities of 2.5 and -2.3 km s$^{-1}$.

\section{Conclusions}
\label{sec:conclusions}

We find, and quantify, evidence for an outflow entrained from a wide-angle wind from the HL Tau system using high resolution ALMA Science Verification observations of CO (J=1-0). We find that the wind is indeed perpendicular to the disk, and that its inclination angle (44$^\circ$), combined with the wind opening angle (90$^\circ$) requires the wind to cross the plane of the sky. This crossing is seen in `counter-flow' emission at 5.5 and 7.5 km s$^{-1}$.  From these angles,  we were able to quantify a characteristic wind velocity ($\sim 2.4$ km s$^{-1}$) in both lobes. 

 \section*{Acknowledgements}
 The authors would like to acknowledge helpful discussions with both Jane Greaves and John Ilee during the preparation of this manuscript. AJ acknowledges the support of the DISCSIM project, grant agreement 341137 funded by the European Research Council under ERC-2013-ADG. This research made use of NASA's Astrophysics Data System; APLpy, an open-source plotting package for Python hosted at http://aplpy.github.com; Astropy, a community-developed core Python package for Astronomy \citep{astropy}. This paper makes use of the following ALMA data: ADS/JAO.ALMA\#2011.0.00015.S. ALMA is a partnership of ESO (representing its member states), NSF (USA) and NINS (Japan), together with NRC (Canada) and NSC and ASIAA (Taiwan), in cooperation with the Republic of Chile. The Joint ALMA Observatory is operated by ESO, AUI/NRAO and NAOJ.
  
\bibliographystyle{mnras}
\bibliography{../../../bibtest}

\begin{thebibliography}{}
\makeatletter
\relax
\def\mn@urlcharsother{\let\do\@makeother \do\$\do\&\do\#\do\^\do\_\do\%\do\~}
\def\mn@doi{\begingroup\mn@urlcharsother \@ifnextchar [ {\mn@doi@}
  {\mn@doi@[]}}
\def\mn@doi@[#1]#2{\def\@tempa{#1}\ifx\@tempa\@empty \href
  {http://dx.doi.org/#2} {doi:#2}\else \href {http://dx.doi.org/#2} {#1}\fi
  \endgroup}
\def\mn@eprint#1#2{\mn@eprint@#1:#2::\@nil}
\def\mn@eprint@arXiv#1{\href {http://arxiv.org/abs/#1} {{\tt arXiv:#1}}}
\def\mn@eprint@dblp#1{\href {http://dblp.uni-trier.de/rec/bibtex/#1.xml}
  {dblp:#1}}
\def\mn@eprint@#1:#2:#3:#4\@nil{\def\@tempa {#1}\def\@tempb {#2}\def\@tempc
  {#3}\ifx \@tempc \@empty \let \@tempc \@tempb \let \@tempb \@tempa \fi \ifx
  \@tempb \@empty \def\@tempb {arXiv}\fi \@ifundefined
  {mn@eprint@\@tempb}{\@tempb:\@tempc}{\expandafter \expandafter \csname
  mn@eprint@\@tempb\endcsname \expandafter{\@tempc}}}

\bibitem[\protect\citeauthoryear{{ALMA Partnership} et~al.,}{{ALMA Partnership}
  et~al.}{2015}]{Brogan15}
{ALMA Partnership} et~al., 2015, eprint arXiv:1503.02649

\bibitem[\protect\citeauthoryear{Alexander, Clarke  \& Pringle}{Alexander
  et~al.}{2006}]{Alexander06}
Alexander R.~D.,  Clarke C.~J.,   Pringle J.~E.,  2006, Monthly Notices of the
  Royal Astronomical Society, 369, 216

\bibitem[\protect\citeauthoryear{Arce, Shepherd, Gueth, Lee, Bachiller, Rosen
  \& Beuther}{Arce et~al.}{2007}]{Arce07}
Arce H.~G.,  Shepherd D.,  Gueth F.,  Lee C.-F.,  Bachiller R.,  Rosen A.,
  Beuther H.,  2007, in Proceedings. pp 245--260

\bibitem[\protect\citeauthoryear{Beck, Bary  \& McGregor}{Beck
  et~al.}{2010}]{Beck10}
Beck T.~L.,  Bary J.~S.,   McGregor P.~J.,  2010, The Astrophysical Journal,
  722, 1360

\bibitem[\protect\citeauthoryear{Carrasco-Gonz{\'a}lez, Rodr{\'\i}guez, Anglada
   \& Curiel}{Carrasco-Gonz{\'a}lez et~al.}{2009}]{Carrasco09}
Carrasco-Gonz{\'a}lez C.,  Rodr{\'\i}guez L.~F.,  Anglada G.,   Curiel S.,
  2009, The Astrophysical Journal Letters, 693, L86

\bibitem[\protect\citeauthoryear{Frank et~al.,}{Frank et~al.}{2014}]{Frank14}
Frank A.,  et~al., 2014, in , Protostars and Planets VI.
University of Arizona Press, pp 451--474

\bibitem[\protect\citeauthoryear{Klaassen et~al.,}{Klaassen
  et~al.}{2013}]{Klaassen13a}
Klaassen P.~D.,  et~al., 2013, Astronomy and Astrophysics, 555, A73

\bibitem[\protect\citeauthoryear{Krist, Stapelfeldt, Hester, Healy, Dwyer  \&
  Gardner}{Krist et~al.}{2008}]{Krist08}
Krist J.~E.,  Stapelfeldt K.~R.,  Hester J.~J.,  Healy K.,  Dwyer S.~J.,
  Gardner C.~L.,  2008, The Astronomical Journal, 136, 1980

\bibitem[\protect\citeauthoryear{Kwon, Looney  \& Mundy}{Kwon
  et~al.}{2011}]{Kwon11}
Kwon W.,  Looney L.~W.,   Mundy L.~G.,  2011, The Astrophysical Journal, 741, 3

\bibitem[\protect\citeauthoryear{Lucas et~al.,}{Lucas et~al.}{2004}]{Lucas04}
Lucas P.~W.,  et~al., 2004, Monthly Notices of the Royal Astronomical Society,
  352, 1347

\bibitem[\protect\citeauthoryear{Lumbreras \& Zapata}{Lumbreras \&
  Zapata}{2014}]{Lumbreras14}
Lumbreras A.~M.,  Zapata L.~A.,  2014, The Astronomical Journal, 147, 72

\bibitem[\protect\citeauthoryear{McMullin, Waters, Schiebel, Young  \&
  Golap}{McMullin et~al.}{2007}]{CASA}
McMullin J.~P.,  Waters B.,  Schiebel D.,  Young W.,   Golap K.,  2007, {CASA
  Architecture and Applications}

\bibitem[\protect\citeauthoryear{Movsessian, Magakian  \& Moiseev}{Movsessian
  et~al.}{2012}]{Movsessian12}
Movsessian T.~A.,  Magakian T.~Y.,   Moiseev A.~V.,  2012, Astronomy and
  Astrophysics, 541, A16

\bibitem[\protect\citeauthoryear{Mundt \& Fried}{Mundt \&
  Fried}{1983}]{Mundt83}
Mundt R.,  Fried J.~W.,  1983, Astrophysical Journal, 274, L83

\bibitem[\protect\citeauthoryear{Mundt, Buehrke, Solf, Ray  \& Raga}{Mundt
  et~al.}{1990}]{Mundt90}
Mundt R.,  Buehrke T.,  Solf J.,  Ray T.~P.,   Raga A.~C.,  1990, Astronomy and
  Astrophysics (ISSN 0004-6361), 232, 37

\bibitem[\protect\citeauthoryear{Panoglou, Cabrit, Pineau~des For{\^e}ts,
  Garcia, Ferreira  \& Casse}{Panoglou et~al.}{2012}]{Panoglou12}
Panoglou D.,  Cabrit S.,  Pineau~des For{\^e}ts G.,  Garcia P. J.~V.,  Ferreira
  J.,   Casse F.,  2012, Astronomy and Astrophysics, 538, 2

\bibitem[\protect\citeauthoryear{Pudritz \& Ouyed}{Pudritz \&
  Ouyed}{1997}]{Pudritz97}
Pudritz R.~E.,  Ouyed R.,  1997, Herbig-Haro Flows and the Birth of Stars; IAU
  Symposium No. 182, 182, 259

\bibitem[\protect\citeauthoryear{Stephens et~al.,}{Stephens
  et~al.}{2013}]{Stephens13}
Stephens I.~W.,  et~al., 2013, Astrophysical Journal, 769, L15

\bibitem[\protect\citeauthoryear{Stephens et~al.,}{Stephens
  et~al.}{2014}]{Stephens14}
Stephens I.~W.,  et~al., 2014, Nature, 514, 597

\bibitem[\protect\citeauthoryear{Takami, Beck, Pyo, McGregor  \& Davis}{Takami
  et~al.}{2007}]{Takami07}
Takami M.,  Beck T.~L.,  Pyo T.-S.,  McGregor P.,   Davis C.,  2007, The
  Astrophysical Journal, 670, L33

\bibitem[\protect\citeauthoryear{{The Astropy Collaboration} et~al.,}{{The
  Astropy Collaboration} et~al.}{2013}]{astropy}
{The Astropy Collaboration} et~al., 2013, Astronomy and Astrophysics, 558, A33

\bibitem[\protect\citeauthoryear{Welch, Hartmann, Helfer  \& Brice{\~n}o}{Welch
  et~al.}{2000}]{Welch00}
Welch W.~J.,  Hartmann L.,  Helfer T.,   Brice{\~n}o C.,  2000, The
  Astrophysical Journal, 540, 362

\makeatother
\end{thebibliography}

\appendix

\section{Calculating Outflow Mass and Kinematics}

Here we present the methods used to quantify the energetics of the outflowing molecular material from the HL Tau disk. We stress that the line emission, especially the blue shifted emission, is likely to be spatially filtered by the observations.  There exist no publicly available single dish CO J=1-0 observations by which we can quantify the amount of filtered flux.

\subsection{Data Processing}

The datacube was masked at 3$\sigma$, which was internally derived using line free channels, and corresponds to the rms noise levels listed in Table \ref{tab:observations}.  Within the masked datacube, using velocity limits listed in Table \ref{tab:energetics1}, we determined the flux in, and noted the velocity of,  each channel. The measured fluxes were converted from mJy/beam km s$^{-1}$ to K km s$^{-1}$ using the size of the synthesised beam ($0.98''\times0.9ее$), and observing frequency (115 GHz).

When calculating quantities from these measured intensities and velocities, we used the python package \texttt{uncertainties} to propagate the measured uncertainties through our calculations. This includes the uncertainty in the flux (taken as the rms noise in each channel) and velocity (taken as half the width of a velocity channel).
\subsection{Calculating Physical Quantities}
\label{sec:outflow_calculations}
With the intensities in each channel we determined the column density in the upper state (J=1) level of CO.  Then, assuming LTE and a temperature of 50 K, this was then scaled using the partition function to calculate the  total column density of CO we are observing using:

\begin{equation}
N_1 = \frac{8 k \pi \nu^2}{h c^3}\frac{1}{A_{10}}\int T dv
\end{equation}
\begin{equation}
N_{tot} = Z(T) * N_1 / (2J+1) * e^{\frac{-E_1}{kT}} 
\end{equation}

Where $N_1$ is the column density in the J=1 level, and $A_{10}$ is the Einstein A coefficient for the J=1-0 transition of CO.  With the assumption of LTE, we extrapolated this to the total CO column density using the second equation above, where $Z(T)$  is the partition function, $J$ is the upper state quantum number, and $E_1$ is the energy of the J=1 state.

The column density calculated above corresponds to the column density within each velocity channel.  The total column density integrated over the velocity ranges listed in Table \ref{tab:energetics1}, was calculated by summing the column densities in each channel.

The integrated red and blue shifted masses were calculated from the column densities under the assumption of optically thin emission:

\begin{equation}
M_{ch} = N_{tot} * n_{\rm beams} * A_{\rm beams} * m_{H_2} / X
\end{equation}

Where $n_{\rm beams}$ is the number of beams the column density is being summed over,  $A_{\rm beams}$ is the area of each beam (in cm$^{2}$), $m_{H_2}$ is the mean molecular mass of hydrogen, and  $X$ is the abundance of CO, which we assume to be 10$^{-4}$.  Subsequently, we determined the total blue and redshifted outflowing masses by summing over the velocities listed in Table \ref{tab:energetics1}.

One of the key reasons for calculating the mass in each channel independently is to best quantify the momentum and energy in the flow.  For each velocity channel, we multiplied the derived mass by the LSR corrected gas velocity ($v_{ch} = v_{\rm means} - v_{\rm LSR}$) to determine the momentum in that channel Then the absolute values of the momenta in each channel were summed to determine the total momentum in both the red and blue shifted emission.  Similarly, to determine the total outflow energy:

\begin{equation}
E = \sum 0.5*  M_{ch}*v_{ch}^2
\end{equation}

The outflow mechanical luminosity and mass loss rates were determined by dividing the total outflow energy and mass (in each lobe) by the dynamical age of the outflow. This age was determined by dividing the extent of the wind by the maximum velocity of the gas in the flow.  Given a wind length of 20$``$, and a maximum velocity of 5 km s$^{-1}$, at a distance of 140 pc, we derive an outflow age of approximately  2650 yrs. 

The `wind length' corresponds to the approximate ends of the red-shifted emission along the limb brightened edges of the wind. For scale, 20$``$ corresponds to the distance between the disk and the ends of the green lines in Figure \ref{fig:CO_mom1} .  This outflow age estimate is subject to a number of caveats, chief amongst which are the lack of short spacing information likely filtering out how large the outflowing material is, the possibility that the wind could be accelerating (or decelerating) with time, and there being emission at higher velocities which are not detected due to the sensitivity limits of the observations.

\end{document}